\documentclass[10pt,journal,compsoc]{IEEEtran}

\usepackage{tikz}
\usepackage{graphicx}
\usepackage[utf8]{inputenc}
\usepackage{mathtools}
\usepackage{amsmath}
\usepackage{pgfplots}
\usepackage{breqn}

\usepackage{subfiles}
\usepackage[justification=centering,font=scriptsize,skip=0pt, belowskip=-1pt,aboveskip=0pt]{caption}
\usepackage{amsmath}
\usepackage{epstopdf}
\usepackage{lipsum}
\usepackage[noadjust]{cite}
\usepackage[ruled, vlined, linesnumbered]{algorithm2e}
\SetAlFnt{\small\sffamily}

\SetCommentSty{mycommfont}
\usepackage{color}
\usepackage{pifont}
\newcommand{\subparagraph}{}
\usepackage{titlesec}

\usepackage[T1]{fontenc}
\usepackage{mdframed} % putting section into box
\usepackage{footmisc}

\newcommand{\lightchain}{LightChain\xspace}
\newcommand{\systemCapacity}{n\xspace}
\newcommand{\blockCapacity}{b\xspace}
\newcommand{\transaction}{tx\xspace}
\newcommand{\idSize}{s\xspace}

\newcommand{\validatorThreshold}{\alpha\xspace}
\newcommand{\signatureThreshold}{t\xspace}
\newcommand{\ufp}{q\xspace} %uniform failure probability 
\newcommand{\adv}{f\xspace} %adversarial fraction
\newcommand{\prev}{prev\xspace}
\newcommand{\txnum}{$min\_tx$\xspace}

\newcommand{\storageGain}{66}
\newcommand{\bootstrapGain}{380}
\usepackage[english]{babel}

\newtheorem{corollary}{Corollary}[section]

\let\OLDthebibliography\thebibliography
\renewcommand\thebibliography[1]{
 \OLDthebibliography{#1}
 \setlength{\parskip}{0pt}
 \setlength{\itemsep}{0pt plus 0ex}
}

\expandafter\def\expandafter\normalsize\expandafter
{%
    \normalsize
    \setlength\abovedisplayskip{2pt}
    \setlength\belowdisplayskip{0pt}
    \setlength\abovedisplayshortskip{0pt}
    \setlength\belowdisplayshortskip{0pt}
}

\title{LightChain: Scalable DHT-based Blockchain \thanks{© 20xx IEEE. Personal use of this material is permitted. Permission from IEEE must be obtained for all other uses, including reprinting/republishing this material for advertising or promotional purposes, collecting new collected works for resale or redistribution to servers or lists, or reuse of any copyrighted component of this work in other works.}
\thanks{Hassanzadeh-Nazarabadi, Y., Küpçü, A. and Özkasap, Ö., 2021. LightChain: Scalable DHT-Based Blockchain. IEEE Transactions on Parallel and Distributed Systems, 32(10), pp.2582-2593. DOI: 10.1109/TPDS.2021.3071176}}

\author{Yahya Hassanzadeh-Nazarabadi, Alptekin K\"{u}p\c{c}\"{u}, and \"{O}znur \"{O}zkasap\\Department of Computer Engineering, Ko\c{c} University, İstanbul, Turkey\\
{\{yhassanzadeh13, akupcu, oozkasap\}}@ku.edu.tr}

\begin{document}

\IEEEtitleabstractindextext
{
\begin{abstract}
As an append-only distributed database, blockchain is utilized in a vast variety of applications including the cryptocurrency and Internet-of-Things (IoT). The existing blockchain solutions show downsides in communication and storage scalability, as well as decentralization. In this paper, we propose \textit{\lightchain}, which is the first blockchain architecture that operates over a Distributed Hash Table (DHT) of participating peers. \textit{\lightchain} is a permissionless blockchain that provides addressable blocks and transactions within the network, which makes them efficiently accessible by all peers. Each block and transaction is replicated within the DHT of peers and is retrieved in an on-demand manner. Hence, peers in \textit{\lightchain} are not required to retrieve or keep the entire ledger. \textit{\lightchain} is fair as all of the participating peers have a uniform chance of being involved in the consensus regardless of their influence such as hashing power or stake. We provide formal mathematical analysis and experimental results (simulations and cloud deployment)
to demonstrate the security, efficiency, and fairness of \textit{\lightchain}, and show that \textit{\lightchain} is the only existing blockchain that can provide integrity under the corrupted majority power of peers. 
As we experimentally demonstrate, compared to the mainstream blockchains such as Bitcoin and Ethereum, \textit{\lightchain} requires around $\storageGain$ times smaller per node storage, and is around $\bootstrapGain$ times faster on bootstrapping a new node to the system, and each \textit{\lightchain} node is rewarded equally likely for participating in the protocol. 
\end{abstract}

}

\maketitle 
\section{Introduction}
\label{lightchain:sec_intro}
\IEEEPARstart{B}lockchain \cite{nakamoto2008bitcoin} is an append-only distributed database that provides a partial ordering of blocks among a set of trust-less peers. The blocks consist of transactions and are connected via immutable links from each block to its previous one on a chain that is called the \textit{ledger}. Because they define a partial ordering of blocks without the need of a globally synchronized clock, provide a tamper-proof architecture, and establish trust over a trust-less system of independent peers, the blockchain systems are exploited in numerous decentralized applications including the crypto-currencies \cite{nakamoto2008bitcoin}, Internet-of-Things \cite{aggarwal2019blockchain, neeraj2021probabilistic}, energy trading platforms \cite{jindal2019survivor, chaudhary2019best}, donation systems \cite{biccer2020anonymous}, smart grids \cite{kumar2021industry}, and P2P cloud storage systems \cite{hassanzadeh2019decentralized}. A blockchain system is usually modeled as a layered-architecture with at least four layers, which from bottom to top are named as \textit{Network, Consensus, Storage, and View} layers \cite{croman2016scaling}. The Network layer deals with the dissemination mechanism of the transactions and blocks among the peers of the system. The Consensus layer represents the protocols for the block generation decision-making process, which aim at providing a sequential ordering of the blocks among all the peers. The Storage layer provides the read functionality for the peers to read from the blockchain. The View layer represents the most recent state of the participating peers' data considering all the updates on the ledger from the very first to the most recent blocks.
 
The existing blockchain solutions show scalability problems in at least one layer of the aforementioned architecture. To the best of our knowledge, at the Network layer, all the existing blockchains operate on unstructured overlays \cite{nakamoto2008bitcoin, cryptoeprint:2016:919, kiayias2017ouroboros, bentov2014proof, luu2016secure, kokoris2018omniledger, rocket2018snowflake, chepurnoy2016prunable, otte2017trustchain, zamani2018rapidchain, androulaki2018hyperledger, cachin2016architecture}. Such overlays have no deterministic, well-defined, and efficient lookup mechanism to retrieve the address of the peers, the content of the blocks, and the new transactions. Rather, the knowledge of a peer of other peers, blocks, and transactions is gained by the epidemic message dissemination among the peers (e.g., broadcasting in Bitcoin \cite{nakamoto2008bitcoin}) which costs a message complexity of $O(\systemCapacity)$ to disseminate a new block or transaction through the entire system, where $\systemCapacity$ is the number of participating peers in the system. 

At the Consensus layer, the existing solutions degrade the decentralization of the system by delegating the block generation decision making to a biased subset of the special peers \cite{zheng2017overview}, e.g., the peers with higher computational power \cite{nakamoto2008bitcoin, chepurnoy2016prunable, otte2017trustchain}, higher stakes \cite{cryptoeprint:2016:919}, or longer activity history in the system \cite{bentov2014proof}. 
The existing blockchains are also prone to the consistency problems that are caused by their probabilistic fork-resolving approach at the Consensus layer, i.e., following the longest chain of the forks as the main chain \cite{nakamoto2008bitcoin}. This causes the block generated by the existing consensus solutions is not considered finalized until some new blocks come after it and make that block residing on the longest chain (i.e., main chain) \cite{zheng2018blockchain}. The main chain of existing solutions is not deterministic and can be switched abruptly once another longer chain is introduced by other peers, which compromises the consistency of blockchain.

At the Storage layer, the existing blockchains require a memory complexity of $O(\blockCapacity)$ by downloading and keeping the entire ledger locally at the peer's storage \cite{croman2016scaling}, where $\blockCapacity$ is the number of blocks in the system. In other words, as peers are not able to efficiently lookup any information within the unstructured overlay, they locally store the perceived information and gradually construct a local copy of the entire ledger, which takes $O(\blockCapacity)$ storage complexity. 
Likewise, upon joining the system, during the bootstrapping phase, a new peer needs to verify the entire state of the ledger from the very first block to the most recent one to check the integrity of the ledger \cite{zheng2017overview}. This imposes a time and communication complexity of $O(\blockCapacity)$ at the View layer. Bootstrapping is defined as the process in which a new node constructs its view of the blockchain \cite{croman2016scaling}. 

The best existing approach to overcome the mentioned scalability problems of the blockchains is to apply sharding \cite{kokoris2018omniledger, luu2016secure, zamani2018rapidchain}, i.e., splitting the system into multiple smaller groups of peers, and each group operates in parallel on a disjoint version of the ledger. Although this reduces the required storage complexity of the nodes to $O(\frac{\blockCapacity}{\log{\systemCapacity}})$, it does not improve the $O(\systemCapacity)$ communication complexity for processing a single transaction or block \cite{zamani2018rapidchain}. 

In this paper, to provide a scalable blockchain architecture with fully decentralized and uniform block generation decision-making, we propose \textbf{\textit{\lightchain}}, which is a permissionless blockchain defined over a Skip Graph-based peer-to-peer (P2P) Distributed Hash Table (DHT) overlay \cite{aspnes2007skip}. \textit{\lightchain} is permissionless \cite{wust2017you} and unmoderated \cite{dousti2020moderated}, as it allows every peer to freely join the blockchain system and participate in the block generation
decision-making without any authority. At the Network layer, \textit{\lightchain} operates on top of a Skip Graph that is a DHT-based structured P2P system with a well-defined topology and deterministic and efficient lookup strategy for data objects. We model each peer, block, and transaction by a Skip Graph node. Compared to the existing blockchains that rely on epidemic dissemination and retrieval of data with the message complexity of $O(\systemCapacity)$, our DHT-based \textit{\lightchain} enables participating peers to make their blocks and transactions addressable and efficiently accessible at the Network layer with the message complexity of $O(\log{\systemCapacity})$. 

At the Consensus layer, we propose Proof-of-Validation (PoV) as an efficient, immutable, and secure consensus protocol for \textit{\lightchain}. PoV is efficient as it requires only $O(\log{\systemCapacity})$ message complexity for validating a single transaction or block. PoV is fair since each participating peer in the system has a uniform chance of being involved in the consensus regardless of its influence, e.g., processing power, available bandwidth, or stake. PoV is immutable as none of the (influential) peers in reaching a consensus can legitimately change the consensus at a later time after it is finalized. PoV is secure as the malicious peers are not able to generate and append an illegitimate transaction or block to the ledger. In PoV, the validation of each block is designated to a subset of the peers, which are chosen uniformly for each block based on its hash value (modeled as a random oracle), and are contacted efficiently using the DHT overlay. In contrast to existing consensus approaches, our PoV preserves the integrity and consistency of the blockchain in the presence of colluding adversarial peers (e.g., Sybil adversary \cite{douceur2002sybil}), as well as selfish miners \cite{eyal2018majority,biccer2020fortis}, as no peer can contribute to the consensus of any two consecutive blocks unless with a negligible probability in the security parameter of the system. As we show both experimentally and analytically in the rest of the paper, there is no threshold for the number of adversarial peers in PoV, and it can be configured to remain secure and immutable even when the adversarial peers become the majority. PoV also governs a deterministic rule on resolving the forks at the Consensus layer, which makes the main chain recognized in a deterministic fashion and followed by all the peers.  

At the Storage layer \textit{\lightchain} provides storage scalability and load distribution for the peers by utilizing DHT-based replication. Each peer in \textit{\lightchain} is only responsible for keeping a small subset of the randomly assigned blocks and transactions while accessing the transactions and blocks replicated on other peers of the system on an on-demand basis using the efficient Skip Graph retrievability. To preserve the data availability in presence of malicious peers, the replication in \textit{\lightchain} is done in a way that it provides at least one copy of each block and transaction accessible at any time with a very high probability in the security parameter of the system. Having $\blockCapacity$ blocks in the system, compared to the existing shardless solutions that require memory complexity of $O(\blockCapacity)$ on each node, our \textit{\lightchain} imposes only a memory complexity of $O(\frac{\blockCapacity}{\systemCapacity})$ on each node.

At the View layer, \textit{\lightchain} provides each new node to the system with a fast bootstrapping mechanism called \textit{Randomized Bootstrapping}, which takes $O(\log{\systemCapacity})$ message complexity and $O(\systemCapacity)$ time complexity to construct its view of the entire ledger. Hence, opposed to the existing solutions that take $O(\blockCapacity)$ message and time complexity ($\blockCapacity >> \systemCapacity$), \textit{\lightchain} enables immediate participation on the blockchain system without the need to verify the entire blockchain for new nodes. Operating on a DHT-based overlay of peers, \textit{\lightchain} also provides a novel approach to directly obtain a particular peer's state without the need to traverse the entire ledger, which takes the message complexity of $O(\log{\systemCapacity})$ and the time complexity of $O(1)$.

The original contributions of this paper are as follows. 
\begin{itemize}
    \item We propose \textit{\lightchain}, the first fully decentralized scalable, permissionless, and DHT-based blockchain architecture, with asymptotic communication and storage complexities' superiority compared to the existing blockchain solutions. \textit{\lightchain} provides full decentralization and fairness in block-generation decision making by providing each peer of the system with a uniform chance of being involved in the consensus regardless of its influence, e.g., processing power, available bandwidth, or stake. 
    
    \item We provide security definitions for \textit{\lightchain}, and analyze how to set its operational parameters to achieve those security features. 
  
    \item We extended the Skip Graph simulator SkipSim \cite{hassanzadeh2020skipsim} with the blockchain-based simulation scenarios, implemented and simulated the \textit{\lightchain} in extensive simulation scales of $10K$ nodes, and show its performance concerning the security features, in the presence of colluding adversarial nodes. 
    
    \item We also implemented a proof-of-concept version of \textit{\lightchain} node \cite{hassanzadeh2020containerized}, deployed it as an operational \textit{\lightchain} system on Google Cloud Platform, and measured its operational overheads in practice. 
\end{itemize}

\section{Related Works}
\label{lightchain:sec_relatedworks}
\textbf{Network Layer:} Dissemination of a new transaction or block in the existing blockchains is done via Broadcasting \cite{nakamoto2008bitcoin}, Flooding \cite{chepurnoy2016prunable}, or Gossiping \cite{otte2017trustchain}, which require $O(\systemCapacity)$ message complexity for a single block or transaction to be accessible by every peer of the system. On the other hand, our proposed \textit{\lightchain} applies a message complexity of $O(\log{\systemCapacity})$ to insert a new transaction or block in the Skip Graph DHT overlay, and make it accessible by every peer of the system. Additionally, in our proposed \textit{\lightchain}, the latest state of the data objects are addressable within the network (e.g., the balance of a node in cryptocurrency applications), and retrievable with the message complexity of $O(\log{\systemCapacity})$. Therefore, in contrast to the existing blockchains, peers in \textit{\lightchain} are not required to keep searching and retrieving the most recent blocks frequently. Rather, they can search and retrieve the latest state of their data objects of demand interest.

\textbf{Consensus Layer:} In \textit{Proof-of-Work (PoW)} \cite{nakamoto2008bitcoin, wood2014ethereum}, the block generation decision-making is heavily correlated with the hash power of peers, which sacrifices the fairness and decentralization of the system in favor of the nodes with higher hash power. PoW tolerates up to $\frac{1}{2}$ fraction of malicious hashing power. \textit{Proof-of-Stake (PoS)}-based approaches, degrade the decentralization of the system by binding the block generation decision-making power of the nodes with the amount of stake they own \cite{cryptoeprint:2016:919, kiayias2017ouroboros, buterin2017casper}. PoS can tolerate less than $\frac{1}{3}$ fraction of malicious stakeholders. However, they apply a message complexity of $\Omega(\systemCapacity)$ to the system for block generation decision making.
In the \textit{Byzantine Fault Tolerance (BFT)}-based approaches, each node broadcasts its vote to the others, receives their votes, and follows the majority \cite{cachin2016architecture, androulaki2018hyperledger, schwartz2014ripple, kwon2014tendermint, rocket2018snowflake, castro1999practical}. Practical BFT (PBFT) \cite{castro1999practical} is one of the pioneers in the BFT-based class that is adaptable to the practical asynchronous setups (e.g., the Internet) where there are no known bounds on the relative execution speeds of the nodes and the message-delivery latency among them \cite{tanenbaum2007distributed}. Although such BFT-based consensus protocols can tolerate up to $\frac{1}{3}$ fraction of the adversarial nodes \cite{pease1980reaching}, they apply a message complexity of $O(\systemCapacity^{2})$ to the system for processing a single transaction or block \cite{xiao2020survey}. Some efforts are done to improve the message overhead in such systems. For example, in Ripple \cite{schwartz2014ripple} instead of reaching a consensus with the majority of the system, each node only reaches a consensus with a subset of nodes it has trust on. Although this slightly improves the message overhead per node depending on the size of its trusted list, it degrades the fault tolerance of the system to less than a fraction of $\frac{1}{5}$ of malicious nodes (in contrast to the classical $\frac{1}{3}$).
In the sharding-based approaches the system is partitioned into disjoint subsets of peers, e.g., subsets of size $O(\log{\systemCapacity})$ \cite{zamani2018rapidchain}. Each subset is working on an independent version of the ledger using BFT in an epoch-based manner \cite{luu2016secure, kokoris2018omniledger}. Although intra-shard transactions are processed with a lower message complexity, the inter-shard transactions should go through the main ledger, which requires the same message complexity as the case of BFT in the no-shard scenarios. As we define in Section \ref{lightchain:sec_intro}, compared to the existing PoW and PoS consensus solutions, our proposed Proof-of-Validation (PoV) is the only one that provides security, immutability, decentralization, and fairness altogether. Likewise, compared to the BFT-based mechanisms, our PoV only applies a message complexity of $O(\log{\systemCapacity})$ to the system for reaching consensus over a block. Finally, there is no inherent adversarial fraction limitation in PoV. Rather, it receives the upper bound on the adversarial fraction of the nodes as input and tweaks its parameters to preserve the security and immutability accordingly. Considering that in PoV the block generation decision making chance is uniformly distributed among the participating nodes regardless of their influence in the system, as we experimentally show in Section \ref{lightchain:sec_results}, our proposed PoV remains secure and immutable even when the fraction of malicious nodes becomes the majority, which is in contract to the existing PoW-, PoS-, and BFT- consensus solutions.

\textbf{Storage Layer:} Having $\blockCapacity$ blocks in the system, most of the existing blockchains apply a storage complexity of $O(\blockCapacity)$ on each peer \cite{nakamoto2008bitcoin, wood2014ethereum, croman2016scaling}. The best existing blockchain architectures concerning storage overhead are the shard-based ones, which apply a storage complexity of $O(\frac{\blockCapacity}{\log{\systemCapacity}})$ \cite{chepurnoy2016prunable, zamani2018rapidchain}. Current attempts on storing each transaction only on the sender and receiver nodes result in overwhelming time complexity of $O(\systemCapacity \times \blockCapacity)$ on generating new transactions \cite{otte2017trustchain, harris2018holochain}. 
Compared to the existing solutions, our proposed \textit{\lightchain} requires $O(\frac{\blockCapacity}{\systemCapacity})$ storage complexity on each peer without any sharding, and with the decentralization of system fully preserved.

\textbf{View Layer: } To the best of our knowledge, there is no existing secure and fast bootstrapping approach as we have in our proposed \textit{\lightchain}, i.e., $O(\systemCapacity)$ in time and $O(\log{\systemCapacity})$ in message complexity. Having $\blockCapacity$ blocks in the system, the local self-construction of view from scratch in the existing blockchains takes the time and message complexity of $O(\blockCapacity)$ $(\blockCapacity >> \systemCapacity)$ on each node by collecting all the blocks, reconstructing the ledger locally, and computing the state of each node by replaying all transactions.

Table \ref{lightchain:table_related_works} compares a variety of the best existing blockchain solutions to our proposed \textit{\lightchain} across different layers of blockchain protocols stack. 

\begin{table*}
\centering
{
    \begin{tabular}{ |l|l|l|l|l|  }
    \hline
    Strategy & Network &Consensus & Storage & View\\ 
    \hline
    %Bitcoin
    Bitcoin \cite{nakamoto2008bitcoin} & Broadcasting ($O(\systemCapacity)$) & PoW & Full ($O(\blockCapacity)$) & Self-Construction ($O(\blockCapacity)$) \\
    %Rollerchain
    Rollerchain \cite{chepurnoy2016prunable} & Flooding ($O(\systemCapacity)$)  & PoW & Distributed ($O(\frac{\blockCapacity}{\log{\systemCapacity}})$) & Self-Construction ($O(\blockCapacity)$) \\
    %Trustchain
    Trustchain \cite{otte2017trustchain} & Gossiping ($O(\systemCapacity)$)  & PoW & Distributed ($O(\blockCapacity)$) & Self-Construction ($O(\blockCapacity)$) \\
    %Snow White
    Snow White \cite{cryptoeprint:2016:919} & Broadcasting ($O(\systemCapacity)$)  & PoS & Full ($O(\blockCapacity)$) & Self-Construction ($O(\blockCapacity)$) \\
    %PoA-Bitcoin
    PoA-Bitcoin \cite{bentov2014proof} & Broadcasting ($O(\systemCapacity)$)  & PoW-PoS & Full ($O(\blockCapacity)$) & Self-Construction ($O(\blockCapacity)$) \\
    %Elastico
    Elastico \cite{luu2016secure} & Broadcasting ($O(\systemCapacity)$)  & BFT & Full ($O(\blockCapacity)$) & Self-Construction ($O(\blockCapacity)$) \\
    %Rapidchain
    Rapidchain \cite{zamani2018rapidchain} & Gossiping($o(\systemCapacity)$) & BFT & Distributed ($O(\frac{\blockCapacity}{\log{\systemCapacity}})$) & Self-Construction ($O(\blockCapacity)$) \\
    %Bitcoin-NG
    BitCoin-NG \cite{eyal2016bitcoin} & Broadcasting ($O(\systemCapacity)$)   & PoW & Full ($O(\blockCapacity)$) & Self-Construction ($O(\blockCapacity)$) \\
    %Ouroboros
    Ouroboros \cite{kiayias2017ouroboros} & Broadcasting ($O(\systemCapacity)$) & PoS & Full ($O(\blockCapacity)$) & Self-Construction ($O(\blockCapacity)$) \\
    %Omniledger
    Omniledger \cite{kokoris2018omniledger} & Gossiping ($O(\systemCapacity)$) & BFT & Full ($O(\blockCapacity)$) & Self-Construction ($O(\blockCapacity)$) \\
    %Avalanche
    Avalanche \cite{rocket2018snowflake} & Gossiping ($O(\systemCapacity)$) & Snowflake & Full ($O(\blockCapacity)$) & Self-Construction ($O(\blockCapacity)$) \\
    %Lightchain
    \textbf{\textit{\lightchain}} & \textbf{DHT ($O(\log{\systemCapacity})$)} &
    \textbf{PoV} & \textbf{Distributed} ($O(\frac{\blockCapacity}{\systemCapacity})$) & \textbf{Randomized-Bootstrapping ($O(\log{\systemCapacity})$)} \\
    \hline
    \end{tabular}
}
\caption{A comparison among the best existing blockchain solutions in a system with $\systemCapacity$ nodes and $\blockCapacity$ blocks. We assume an approach supports distributed storage, if the storage load of blocks and transactions is distributed among all the participating peers in a policy-based manner, e.g., replication. Also, by the self-construction at the view layer we mean a peer is required to collect all blocks, build the ledger locally, and construct its view by traversing the ledger from the tail to the head.}
\label{lightchain:table_related_works}
\end{table*}

\color{black}

\section{Preliminaries and System Model}
\label{lightchain:sec_preliminaries}
\textbf{Skip Graphs} \cite{aspnes2007skip} are DHT-based overlay of nodes, where each node is identified with an (IP) address and two identifiers: a numerical ID and a name ID. 
Each Skip Graph node can search and find the address of other nodes of Skip Graph that possess a specific numerical ID or name ID, by utilizing a search for numerical ID \cite{aspnes2007skip,hassanzadeh2015locality}, or a search for name ID \cite{hassanzadeh2016laras} of those nodes, respectively. Both searches are done with the message complexity of $O(\log{\systemCapacity})$. As the result of the searches, if the targeted identifier is available in the Skip Graph, the (IP) addresses of their corresponding nodes are returned to the search initiator. Otherwise, the (IP) addresses of the nodes with the most similar identifiers to the search target are returned. As detailed in Section \ref{lightchain:sec_solution}, in \textit{\lightchain}, a Skip Graph overlay of peers is constructed by representing each peer as a Skip Graph node. We assume that each participating peer joins the Skip Graph overlay using the Skip Graph join protocol in a fully decentralized manner \cite{aspnes2007skip}. In this paper, each peer corresponds to a device connected to the Internet (e.g., a laptop, smartphone, smart TV) that executes an instance of the \textit{\lightchain} protocol. We consider the system under constant churn \cite{imtiaz2019churn}, i.e., the participating peers are dynamic between offline and online states. We assume the existence of a churn stabilization strategy \cite{hassanzadeh2019interlaced} that preserves the connectivity of the overlay under churn.

\textbf{Synchronization:} We assume the system is partially synchronous \cite{tanenbaum2007distributed}, meaning that most of the time the process execution speeds and message-delivery times are bounded. To handle the times that the system goes asynchronous where no such bounds are assumed, processes use a timer and timeout mechanism to detect process crash failures.

A \textbf{blockchain} is a linked-list of blocks with an immutable link from each block to its previous one \cite{croman2016scaling}. By an immutable link, we mean that each block points back to the collision-resistant hash value of its previous block on the linked-list. The linked-list of blocks is called \textit{ledger}. In this paper, we call the first block of the ledger (i.e., head) as the \textit{Genesis} block and the most recent block (i.e., tail) as the \textit{current tail}. For each block on the ledger, the \textit{previous} relationship is the immutable link from it to its previous block on the ledger. Blockchain defines a partial ordering of the blocks on the ledger based on the previous relationship. We say that block $blk1$ is the \textit{immediate predecessor} of the $blk2$, if $blk2$ points back to the hash value of $blk1$ as its previous block on the ledger. In this situation, $blk2$ is the \textit{immediate successor} of $blk1$. Due to the immutable links, the blockchain is considered as an append-only database. Updating a block of the ledger by changing its content is not allowed, and is considered an adversarial act. Re-establishing the connectivity of ledger after an update on a single block requires refreshing the hash pointers on all the subsequent blocks, which is a computationally hard problem in the existing blockchains \cite{nakamoto2008bitcoin}. 

\textbf{Notations:}   
In this paper, we say that a block is \textit{committed} to the blockchain if it is being written by the Consensus layer protocol of the blockchain to its storage, i.e., the block passes the defined consensus verification and is being appended to the tail of the ledger. We denote the security parameter of the system by $\lambda$. We denote the hash function $H:\{0,1\}^{*} \rightarrow \{0,1\}^{\idSize}$ as a random oracle \cite{katz2014introduction}, where $\idSize$ is the identifier size of peers in bits. We assume $\idSize$ is polynomial in the security parameter of the system, i.e., $\idSize = poly(\lambda)$. We denote the \textit{System Capacity} by $\systemCapacity$, and define it as the maximum number of registered peers in the system, i.e., $\systemCapacity = O(2^{\idSize})$. Similarly, we denote the \textit{Block Capacity} by $\blockCapacity$ and define it as the maximum number of the generated blocks in the system. We assume $\blockCapacity >> \systemCapacity$.

\textbf{View Layer:} We assume each peer is participating in the blockchain by a set of \textit{assets} as well as a \textit{balance}. The assets set corresponds to the data objects that the peer initially registers on the blockchain via a transaction, and can update it later on by generating new transactions. The balance of a peer is used to cover its transaction generation fees. We consider a transaction as a state transition of the assets of the transaction's owner. View of a participating peer in our system model towards the blockchain is a table of $(numID\,, lastblk\,, state\,, balance)$ tuples. Each tuple represents the view of the peer concerning another peer of the system with the numerical ID of $numID$, where $lastblk$ is the hash value of the last committed block to the blockchain that contains its most recent transaction, and the current state of the assets as well as the remaining balance of $state$ and $balance$, respectively. By the current state, we mean the most recent values of the assets of the peer considering all the generated transactions by that peer from the Genesis block up to the current tail of the blockchain.

\textbf{Authenticated Searches:} We assume that the search queries over the Skip Graph overlay are authenticated by an authentication mechanism in the presence of an adversarial party that adaptively controls a fraction $\adv$ of nodes and is aiming at conducting routing attacks to the system \cite{boshrooyeh2017guard, taheri2020proof}. By the authenticated searches, we mean that the validity of the search results is publicly verifiable through a search proof that is generated by the signing keys of the participating peers on the search path. The search proof also contains the attributes of the peers on the search query path (e.g., identifiers and (IP) addresses) with the last node on the search path considered as the search result. We assume the success chance of the adversary on breaking the authenticated search mechanism and forging a search proof is limited to some negligible function $\epsilon(\lambda)$ (e.g., $\epsilon(\lambda) = 2^{-\lambda}$).

\section{\lightchain System Architecture \protect \footnote{In this section, we present the architecture of \textit{\lightchain} at each layer of the blockchain protocol stack. We skip the detailed algorithmic presentation of \textit{\lightchain}'s protocols for sake of page limit and refer the interested readers to the full version of this paper \cite{hassanzadeh2019lightchain} for those details.}}
\label{lightchain:sec_solution}
\subsection{Overview:}
\label{lightchain:subsec_overview}
\textbf{Overlay:} In \textit{\lightchain} the peers, as well as the transactions and blocks, are represented by Skip Graph nodes. Each peer invokes the join protocol of Skip Graph \cite{aspnes2007skip} using its identifiers and (IP) address and joins the system. Both identifiers of a peer (i.e., its name ID and numerical ID) are the hash value of its public key (i.e., verification key). In this paper, we assume the peers utilize a digital signature scheme that is existentially unforgeable under adaptive chosen message attack \cite{katz2014introduction}. Once the peer joins the Skip Graph overlay, it can efficiently search for any other peer of the system with the message complexity of $O(\log{\systemCapacity})$. Upon joining the Skip Graph overlay, the peer creates its view of the blockchain using \textit{\lightchain}'s Randomized Bootstrapping feature without the need to download and process the entire ledger. 

\textbf{Transaction Generation:} In \textit{\lightchain}, a transaction represents a state transition of the assets of a peer, which is denoted by the \textit{owner} peer of that transaction. For example, in cryptocurrency applications, the asset of a peer is its monetary wealth, and a transaction represents the state transition of the monetary wealth of it. The owner peer casts the state transition into a transaction, computes the identifiers of validators, searches for the validators over Skip Graph overlay, and asks them to validate its transaction. To be validated, each transaction needs to be signed by a system-wide constant number of validators, where their identifiers are chosen randomly for each transaction to ensure security. In addition to security, the idea of validating transactions makes participating nodes in the block generation needless of going through the validation of individual transactions. 

\textbf{Block Generation:} Once the transaction gets validated, the owner inserts it as a node into the Skip Graph overlay, which makes it searchable and accessible by any other peer. The insertion of the transaction is done by invoking the join protocol of Skip Graph using the transaction's identifiers (i.e., its hash value) but the (IP) address of the owner peer itself. The Skip Graph peers route the messages on behalf of the transaction nodes they hold. This idea is similar to the other existing DHTs like Chord \cite{stoica2001chord}, and enables \textit{\lightchain} peers to search and find the new transactions. Upon finding new validated transactions, each peer is able to cast them into blocks, go through the validation procedure (similar to the transactions' case), and insert the validated block into the Skip Graph overlay. Each transaction's owner then removes its transaction node from the overlay once it is successfully included in a validated block. The idea of representing each transaction and block by a Skip Graph node results in any search for the peer or the transactions and blocks that it holds to be routed to the peer's (IP) address, rendering them accessible by every other peer in a fully decentralized manner with the message complexity of $O(\log{\systemCapacity})$. Hence, in \textit{\lightchain}'s Skip Graph overlay, there exist three types of nodes: peers, transactions, and blocks. In other words,  
the Skip Graph overlay acts as a distributed database of the transactions and blocks that are owned by the peers. As elaborated later, the \textit{previous} relationship of blocks stored in a distributed manner on distinct peers defines a blockchain. 

\textbf{Distributed Storage Layer:} In \textit{\lightchain}, each block and transaction is replicated on its owner and validators to support availability, accessibility, and fault tolerance. Using searchable blocks and transactions as well as replication, in \textit{\lightchain} we introduce the idea of a distributed storage layer for the blockchain where participating peers in the consensus only need to keep and maintain a subset of the blocks, and not the ledger entirely. In the rest of this section, unless stated otherwise, by the term node, we mean a peer.  

\textbf{Incentive Mechanism:} As an incentive mechanism, \textit{\lightchain} employs a monetary balance for each participating peer to exchange with other peers and cover the operational fees of appending data to the blockchain \cite{nakamoto2008bitcoin}. \textit{\lightchain} rewards the peers' honest contribution to maintain the connectivity of the system, provide validation service, and generate blocks. Moreover, \textit{\lightchain} encourages honest peers to audit other peers, by rewarding the detection and report of adversarial acts. Malicious behavior is penalized by \textit{\lightchain} upon detection, and the adversarial peers are blacklisted and gradually isolated from the system.

\subsection{\textbf{Structure of Transactions and Blocks}}
\label{lightchain:subsec_txb_structure}
A \textit{\lightchain} transaction, $tx$, is represented by a $(\prev, \, owner, \, cont, \, search\_proof, \, h, \, \sigma)$ tuple, where $\prev$ is the hash value of a committed block to the blockchain. We use the $\prev$ pointer for each transaction $tx$ to define an order of precedence between $tx$ and all the blocks and transactions in the blockchain without the need for any synchronized clock. The block that is referred by $\prev$ takes precedence over $tx$. All the transactions included in the $\prev$ block are assumed to be committed before $tx$ in the essence of time. Following the same convention, all the blocks and transactions that precede $prev$, also precede $tx$. The \textit{owner} represents the identifier of the owner node in the Skip Graph overlay that generates the transaction $tx$.  The contribution field (i.e., $cont$) of a transaction denotes the state transition of the assets of the owner node, e.g., a monetary remittance between two peers in cryptocurrency applications. The $search\_proof$ field of a transaction is the authenticated proof of searches over the peers of the Skip Graph overlay to find the validators of the transaction $tx$, as explained before. The $h$ field of the transaction $tx$ is the hash value of the transaction, which is computed as shown by Equation \ref{lightchain:eq_hash_tx}. The $\sigma$ field of the transaction $tx$ contains the signatures of both the owner as well the validators on its hash value $h$. The validators' signature is a part of \textit{\lightchain}'s consensus strategy and is explained later.
\begin{equation}
    h = H(prev||owner||cont||search\_proof) \label{lightchain:eq_hash_tx}
\end{equation}
Similarly, a \textit{\lightchain} block $blk$ is defined by a $(prev, \, owner, \, \mathcal{S}, \, search\_proof, \, h, \, \sigma)$ tuple. $\mathcal{S}$ represents the set of all the transactions that are included in the block $blk$. The $h$ field of block $blk$ is its hash value, which is computed as shown by Equation \ref{lightchain:eq_hash_blk}. The $\sigma$ field contains the signatures of both the block's owner as well as the block's validators on its hash value $h$. 
\begin{equation}
    h = H(prev||owner||\mathcal{S}||search\_proof) \label{lightchain:eq_hash_blk}
\end{equation}
\subsection{\textbf{Network Layer: Skip Graph overlay}}
In our proposed \textit{\lightchain}, we represent each peer, transaction, and block by a Skip Graph node. This way, all the peers, transactions, and blocks are addressable within the network. In other words, participating nodes (i.e., peers) in \textit{\lightchain} exploit the Skip Graph overlay to search for each other, as well as each others' blocks and transactions. Both the numerical ID and name ID of the peers are the hash value of their public key using a collision-resistant hash function. 
As in a Skip Graph, nodes' identifiers define the topology \cite{aspnes2007skip}; hence, considering the hash function as a random oracle results in the uniform placement of peers in Skip Graph overlay topology, which limits the adversarial power on tweaking the Skip Graph topology for its advantage.

The numerical ID and name ID of a transaction (or a block) in the Skip Graph overlay is its hash value (i.e., $h$) and its corresponding \textit{\prev} field value, respectively. This regulation enables peers to traverse the \textit{\lightchain}'s ledger in both forward and backward directions. Following this convention, in \textit{\lightchain}, having a block with numerical ID (i.e., the hash value) of $h$ and previous pointer value of \textit{prev}, the (IP) address of the peers that hold the immediate predecessor block are obtained by performing a search for the numerical ID of \textit{\prev} in the Skip Graph overlay \cite{aspnes2007skip}. Similarly, the (IP) address of the peers holding the immediate successor transaction(s) or block(s) in the blockchain are obtainable by performing a search for name ID of $h$ over the Skip Graph overlay. This follows the fact that all the immediate successors of the block have its hash value $h$ as their name ID. 
This feature of the \textit{\lightchain} enables the peers to efficiently update their view towards the tail of the blockchain by performing a search for the name ID of their local tail \cite{hassanzadeh2016laras}. The search returns all the blocks that are appended subsequently to their local tail, as well as all the new validated transactions that are waiting to be included in blocks. Additionally, using this feature, a peer does not need to store the entire blockchain locally. Rather, having only a single block of the ledger enables the peer to efficiently retrieve the predecessor and successor blocks to it with the message complexity of $O(\log{\systemCapacity})$. 

\begin{figure}
\centering
\includegraphics[width=\linewidth]{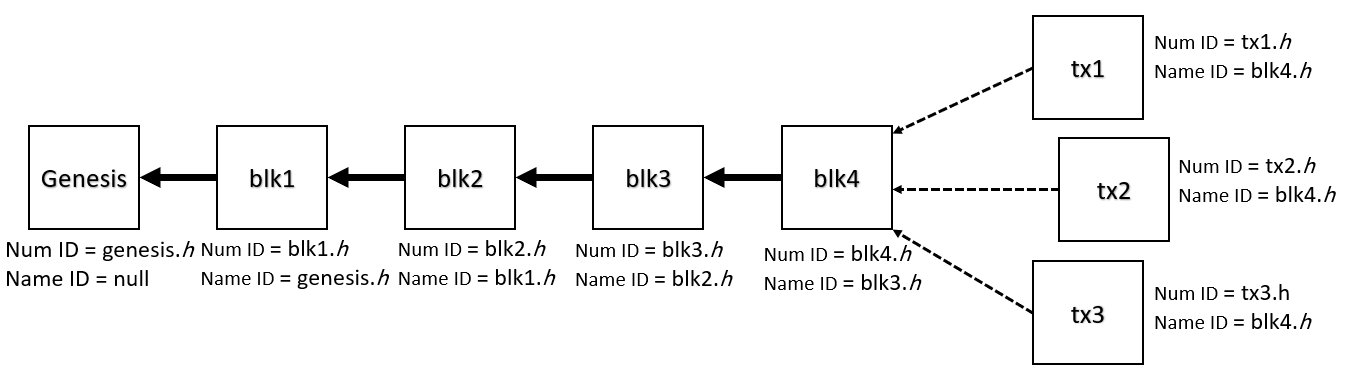}
\caption{The \textit{\lightchain} regulation on name IDs and numerical IDs. Numerical ID (i.e., Num ID) of a block or transaction is its hash value, and name ID is its corresponding \textit{prev} value.} 
\label{figure:lightchain_chain-tx}
\end{figure}

 Figure \ref{figure:lightchain_chain-tx} illustrates this convention of \textit{\lightchain}, where a peer that only has $blk2$ is able to efficiently retrieve its immediate predecessor (i.e., $blk1$) by searching for the numerical ID \cite{aspnes2007skip} of its \textit{\prev} value (i.e., $blk2.prev = blk1.h$) in a fully decentralized manner. The search is responded by the replicas of $blk1$ with their (IP) addresses, and hence the predecessor of $blk2$ (i.e., $blk1$) is retrievable efficiently by directly contacting those replicas. Similarly, the peer that only possesses $blk2$ is able to perform a search for name ID \cite{hassanzadeh2016laras} over its hash value (i.e., $blk2.h$) to retrieve the immediate successor block that comes after $blk2$. As the result of the search for the name ID of $blk2.h$, replicas of $blk3$ respond to the search initiator peer with their (IP) addresses, and $blk3$ is retrievable efficiently by directly contacting those replicas. In the case where a single block has several successor blocks, the search initiator receives a response from each of the immediate successor block replicas. In the example of Figure \ref{figure:lightchain_chain-tx}, considering $blk4$ as the current tail of the blockchain, as discussed later in this section, the newly generated transactions that succeed $blk4$ (i.e., $tx1$, $tx2$, and $tx3$) are efficiently retrievable by performing a search for the name ID using $blk4.h$. 

\subsection{\textbf{Consensus Layer: Proof-of-Validation (PoV)}}
\label{lightchain:subsec_consensus}
The consensus layer of \textit{\lightchain} is based on our proposed PoV, which is a fair, immutable, secure, and efficient consensus protocol (see Section \ref{lightchain:sec_intro} for more details on these features). 
A transaction or block is considered as validated once it successfully passes the PoV consensus. Note that a validated transaction's contribution is not considered effective and authoritative unless it is included in a validated block that is committed to the blockchain. As detailed in the following, PoV provides a set of $\validatorThreshold$ uniformly chosen validators for each transaction or block. PoV then considers a transaction or block as valid if its hash value $h$ is signed by $\signatureThreshold$ validators. Both $\signatureThreshold$ and $\validatorThreshold$ are constant protocol parameters, which are called the \textit{Signatures Threshold} and \textit{Validators Threshold}, respectively. We formally discuss this in Section \ref{lightchain:sec_security}, and develop a formulation for deciding on the proper values of the \textit{Signatures Threshold} and \textit{Validators Threshold} considering the security of the system.

\label{lightchain:tx_validation}
\subsubsection{Transaction Generation and Validation}
For a transaction $\transaction$, the numerical ID of each validator is chosen uniformly as shown by Equation \ref{lightchain:eq_validator}, where $v_{i}$ is the numerical ID of the $i^{th}$ validator in the Skip Graph overlay. 

\begin{equation}
    v_{i} = H(\transaction.\prev||\transaction.owner||\transaction.cont||i) \label{lightchain:eq_validator}
\end{equation}

\noindent The transaction's owner then conducts a search for the numerical ID of the validator (i.e., $v_{i}$) within the Skip Graph overlay. If there exists a peer with the numerical ID of $v_{i}$ in the overlay, the owner receives its (IP) address. Otherwise, it receives the (IP) address of the peer with the largest available numerical ID that is less than $v_{i}$. Both cases are supported by an authenticated search proof that is generated by the Skip Graph peers on the search path and is delivered to the owner. The authenticated proof of the search for the numerical ID of the $i^{th}$ validator is denoted by $search\_proof_{i}$, which also contains all the (IP) addresses and identifiers of the Skip Graph peers on the search path. The last peer on the search path of $v_{i}$ is designated as the $i^{th}$ validator. The transaction's owner then adds the authenticated search proof for all the validators to the transaction, computes its hash value $h$ as specified by Equation \ref{lightchain:eq_hash_tx}, signs the hash value, and appends her signature to $\sigma$. The transaction's owner then contacts the validators asking for the validation of the $tx$. Each validator evaluates the soundness, correctness, and authenticity of $tx$, as well as the balance compliance of its owner to cover the fees. As the validation result for $tx$, the transaction owner either receives a signature over $h$ or $\bot$ from a contacted validator.

\textbf{Soundness:} A transaction $tx$ is sound if it does not precede the latest transaction of its owner on the blockchain, i.e., its $\prev$ should point to the hash value of a validated and committed block on the ledger with no transaction of its owner in any of the subsequent blocks. In other words, soundness requires a causal ordering among the committed transactions of each owner. This is both to counter double-spending from the same set of assets, as well as to make the validation of a transaction a one-time operation, i.e., the owner of a validated $tx$ transaction can append it to the blockchain as long as it does not generate any new transaction on the blockchain that precedes $tx$ based on $\prev$.
Considering the soundness, at most one of the concurrently generated and validated transactions of a peer has the chance to be included in a new block. Once one of its transactions is included in a block, the others become unsound, cannot be included in the same block or further blocks, and should go over re-validation. Therefore, besides preventing the double-spending, soundness provides a uniform chance for the transaction generators to include their transaction into each new block.

\textbf{Correctness:} For a transaction $tx$ to be correct, its contribution field (i.e., $cont$) should represent a valid state transition of the owner's assets. The compliance metric is application dependent. For example, in cryptocurrency applications, for a transaction to be correct, the owner's account should have enough balance to cover the remittance fee (i.e., the contribution). 

\textbf{Authenticity:} The evaluation of authenticity is done by checking the correctness of $h$ based on Equation \ref{lightchain:eq_hash_tx}, verifying $\sigma$ for the inclusion of a valid signature of the transaction's owner over $h$, 
and verifying $search\_proof$ for all the validators of $tx$. A validator rejects the validation of $tx$ as unauthenticated if any of these conditions is not satisfied. 

\textbf{Balance Compliance:} As an incentive mechanism to participate in the validation, \textit{\lightchain} considers a validation fee in the favor of the $\signatureThreshold$ validators of the transaction $tx$ that sign its hash value and make it validated. Also, \textit{\lightchain} considers a routing fee in the favor of all the Skip Graph peers that participate in finding the transaction's validators, i.e., the peers in $search\_proof$. A transaction $tx$ passes the balance compliance part of validation if its owner has enough balance to cover the validation and routing fees. The balance compliance validation is done based on the view of the validator towards the blockchain. Both the routing and validation fees are fixed-value protocol parameters and are the incentive mechanism for the peers to perform the routing and validation honestly \cite{nakamoto2008bitcoin, eyal2016bitcoin}. The fees also prevent Sybil adversarial peers from indefinitely generating transactions by circulating the adversarial balance among themselves and continuously congesting the system with the validation of adversarial transactions. 

Once a transaction $tx$ receives $\signatureThreshold$ valid signatures issued by its uniformly designated validators, it is considered valid and is inserted as a Skip Graph node by its owner, which makes it accessible by other participating peers of the system to be included in a block. The numerical ID of $tx$ is $tx.h$, and the name ID of $tx$ is $tx.\prev$ (block). This enables any Skip Graph peer to conduct a search for name ID on the hash value of any ledger's block within the Skip Graph overlay and find all the new transactions that are pointing back to that block. Setting $tx.\prev$ to point to the tail of the ledger hence increases the chance of $tx$ for being discovered by other peers.

\subsubsection{Block Generation and Validation}
\label{lightchain:blk_validation}
In \textit{\lightchain}, a peer that generates blocks is called a block owner. Once a block owner collects at least \txnum newly generated transactions that have not been included in any committed block to the ledger, it casts them into a new block $blk$, and sends the block for validation. By casting transactions into $blk$ we mean including the collected transactions into the set $\mathcal{S}$ as discussed earlier (i.e., Section \ref{lightchain:subsec_txb_structure}). \txnum is an application-dependent fixed-value parameter of \textit{\lightchain} denoting the minimum number of transactions that should be included in a block. 
To have $blk$ validated, the block owner computes the numerical ID of the $i^{th}$ validator as shown by Equation \ref{lightchain:eq_block_validator}, where $\prev$ is the hash value of the current tail of the ledger. The rest of the finding validators procedure is similar to the transaction case. On receiving a validation request for a block $blk$, each of its PoV validators checks the authenticity and consistency of $blk$ itself, as well as the authenticity and soundness of all transactions included in $\mathcal{S}$ (as discussed earlier). The authenticity evaluation of blocks is done similar to the transactions. 
\begin{equation}
    v_{i} = H(\prev||owner||\mathcal{S}||i) \label{lightchain:eq_block_validator}
\end{equation}

\textbf{Consistency: }A block $blk$ is said to be consistent if its $\prev$ pointer points to the current tail of the blockchain; otherwise it is inconsistent. By the current tail of the blockchain, we mean the most recent view of the validators towards the tail of the chain. However, it is likely for the current tail of the blockchain to be updated during the validation of a newly generated block and make the block under validation inconsistent. To avoid forking the ledger, upon the start of validation, all randomly chosen PoV validators continuously monitor the updates on the tail of the ledger and terminate the validation with rejection (at any step) upon detecting a potential fork that is caused by the current block under validation.

After the $blk$ gets validated, its owner inserts it in the Skip Graph overlay as a node. As the incentive mechanism of \textit{\lightchain}, the owner of a block receives a block generation reward once its block gets validated and committed to the blockchain. The block generation reward is a fixed-value parameter of \textit{\lightchain} that acts both as an incentive mechanism for encouraging the peers to participate progressively in generating blocks, as well as a mean for wealth creation. In this paper, we assume that the generation reward for a block is larger than its validation and routing fees.

\subsubsection{Fork-free mechanism:}
To resolve the forks caused by the simultaneously validated blocks, \textit{\lightchain} governs a fork-free mechanism, which is a deterministic approach that instructs all the peers to solely follow the block with the lowest hash value upon a fork. Upon a fork, we call the block with the lowest hash value as the winner block and the other participating blocks of the fork as the knocked-out ones. The knocked-out block owners remove their block from the Skip Graph overlay, update their set of transactions by dropping the transactions that are included in the winner block, adding the new transactions to reach the \txnum threshold, and restart the validation procedure. The knocked-out block owners neither gain any block generation reward nor lose any balance because of the fees, as these fees and rewards are not effective unless the block is successfully committed to the ledger. To ensure that a newly appended validated block $blk$ to the ledger does not undergo any further fork rivalry, and is considered committed, effective, and finalized, \textit{\lightchain} waits for only one further block to be appended subsequently to $blk$. In this way, as explained earlier, all the forks at the height of the $blk$ are considered as potential forks and are rejected by the consistency checking mechanism of PoV validators. 
\subsection{\textbf{Storage Layer: Replication}}
To provide efficient retrievability and data availability under churn \cite{imtiaz2019churn} in \textit{\lightchain}, each transaction or block is replicated in the local storage of its owner as well as its uniformly chosen PoV validators, and presented by them as  Skip Graph nodes. This makes the transactions and blocks efficiently searchable by all the participating peers in the system. As we show in Section \ref{lightchain:sec_results}, the parameters of \textit{\lightchain} are chosen in a way that at least one (honest) replica for each transaction and block is always available in expectation. Hence, in \textit{\lightchain}, peers do not need to store or download the entire ledger. Rather, they access the transactions and blocks in an on-demand manner, i.e., a peer searches for a transaction or block upon a need and retrieves it efficiently from the overlay.
\subsection{\textbf{View Layer: Randomized Bootstrapping and Fast Retrieval}}
%%%%%%%%%%%%%%%%%%%%%%%%%%%%%%%%%%%%%%%%%%%%%%%%%%%%%%%%%%%%%%%%%%%%%%%%%%%%%%%%%%%%%%%%%%%%%
\label{lightchain:view_layer}
\label{lightchain:subsection_bootstrap}
\textbf{Randomized Bootstrapping:} 
We define the \textit{view introducers} of a new peer as the set of randomly chosen peers that share their view of the blockchain with the newly joined peer. Upon joining the overlay, a new peer computes the numerical IDs of its view introducers based on Equation \ref{lightchain:eq_view_introducers}, where $new\_peer.numID$ is the numerical ID of the new peer and $view\_intro_{i}$ is the numerical ID of the $i^{th}$ view introducer of it. 
The new peer then conducts an authenticated search for the numerical ID of $view\_intro_{i}$ within the overlay \cite{boshrooyeh2017guard}, contacts the peer in the search result, and obtains its view of the blockchain. The new peer continues iterating over $i$ until it obtains $\signatureThreshold$ consistent views. As we show later, we determine  $\signatureThreshold$ and $\validatorThreshold$ in such a way that a new peer obtains $\signatureThreshold$ consistent views of the honest peers by iterating $i$ over $[1, \validatorThreshold]$. 

\begin{equation}
   view\_intro_{i} = H(new\_peer.numID||i)    
\label{lightchain:eq_view_introducers}
\end{equation}

\begin{figure}
\centering
\includegraphics[width=\linewidth]{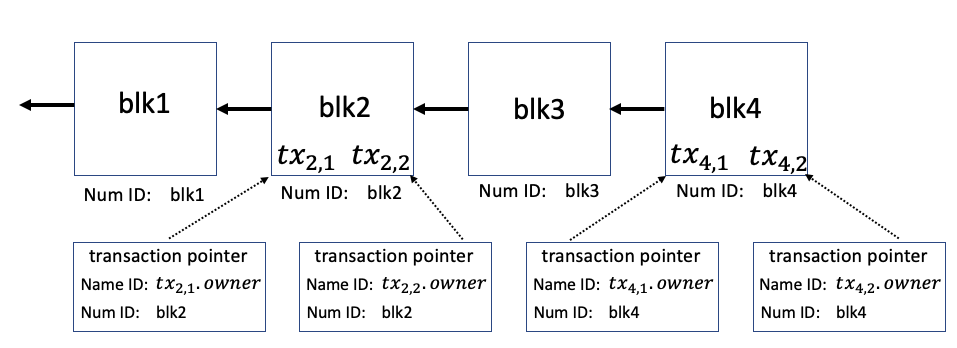}
\caption{Transaction pointers for the transactions of blocks $blk2$ and $blk4$. Transactions and pointers of other blocks are not shown for sake of simplicity.} 
\label{figure:lightchain_txpointers}
\end{figure}

\textbf{Fast Retrieval:} Tracking the updates on the entire view of other peers' assets requires a peer to keep its local view updated with the new blocks, which is a plausible assumption in the majority of the existing solutions \cite{croman2016scaling}. However, in addition to this traditional approach, \textit{\lightchain} enables each peer to directly retrieve the latest assets' state of another peer of interest without the need to keep track of the new blocks on the ledger and sequentially applying the updates. This is done by the additional representation of each block with multiple Skip Graph nodes, one per included transaction, which we call them the associated \textit{transaction pointers} of that block. In this approach, each transaction $tx$ that is included in a committed block $blk$ is represented by a transaction pointer node (i.e., $pointer$) with $pointer.nameID = tx.owner$ and $pointer.numID = blk.h$. The transaction pointer nodes associated with each block are inserted by the block's owner and replicated on the block's PoV validators.
Figure \ref{figure:lightchain_txpointers} depicts an example of transaction pointers for transactions of block $blk2$ and $blk4$ on the ledger. Assume that by the time $blk2$ is committed to the ledger, it contains the transaction $tx_{2,2}$, which is the most recent transaction of $tx_{2,2}.owner$ at the height of $blk2$. Hence, any peer that is solely interested in knowing the latest state of the owner of $tx_{2,2}$ performs a search for the transaction pointer with the name ID of the owner node's identifier, i.e., $tx_{2,2}.owner$ in this example. As result of the search, the querying peer obtains a copy of the transaction pointer. Since the numerical ID of transaction pointer corresponds to the hash of the block that the transaction is included in, by having the transaction pointer, the querying peer finds that the latest transaction of $tx_{2,2}.owner$ is included in block $blk2$. By performing a search for the block numerical ID of $blk2$, the querying peer obtains the content of $blk2$, which includes $tx_{2,2}$ that also contains the latest state of $tx_{2,2}.owner$ affected by the transaction.
To keep track of the latest updates over the state of assets, both the owner and validators of a block should drop each of their associated transaction pointers once an update on the corresponding assets appears on a newer committed block to the ledger. In this way, the transaction pointers act as moving flags within the ledger each pointing to the latest transaction of a peer on the ledger. 
For example, in Figure \ref{figure:lightchain_txpointers} assume that once $blk4$ is committed to the ledger, it contains a new transaction $tx_{4,2}$ from the same owner as $tx_{2,2}$ (i.e., $tx_{2,2}.owner = tx_{4,2}.owner$). In this situation, $blk2$ no longer holds the latest transaction of $tx_{4,2}.owner$. Hence, the owner and PoV validators of $blk2$ drop the transaction pointer for $tx_{2,2}$ from the Skip Graph overlay. 
Taking down a pointer node from the overlay is simply done by performing the Skip Graph node deletion operation \cite{aspnes2007skip} by the owner and each of the validators in a fully decentralized manner. This is for the sake of better efficiency of the search, and to make sure that the transaction pointers always point to the most recent states. Not dropping the (old) pointers after a new update is counted as misbehavior, which we address by the misbehavior detection strategy of \textit{\lightchain}. To address the network asynchrony, however, the block owner and PoV validators are allowed to take down the pointers within a certain block interval upon a new transaction on the associated set of assets that they maintain the pointers. This allows them to have enough time to discover the new updates without being subject to misbehavior. The length of the block interval (i.e., number of blocks between two transaction pointers over the same set of assets) is a constant protocol parameter that is application dependent.  

%%%%%%%%%%%%%%%%%%%%%%%%%%%%%%%%%%%%%%%%%%%%%%%%%%%%%%%%%%%%%%%%
\subsection{Incentive Mechanism}
\label{lightchain:subsection_auditing}
As discussed earlier, the block generation reward and the routing and validation fees constitute the incentive mechanism of \textit{\lightchain} for the peers to retain their honest behavior. The counterpart of honest behavior is the \textit{misbehavior}, which we define as any sort of deviation from the described \textit{\lightchain}'s protocol and architecture. As detailed earlier, for the transactions and blocks that are gone through the consensus layer, we consider the randomly chosen PoV validators to check the submitted transaction or block against the misbehavior. As we show in Section \ref{lightchain:sec_security}, we set the operational parameters of \textit{\lightchain} in a way that an adversarial peer cannot convince the PoV validators on misbehavior unless with a negligible probability in the security parameter. We also introduce the \textit{misbehavior detection} as an extra level of adversarial countermeasure, especially for the adversarial actions that are not gone through the PoV, e.g., direct submission of an invalid block to the ledger. Each peer of \textit{\lightchain} also acts as an auditor for other peers' behavior and gains a \textit{misbehavior audition reward} by reporting their misbehavior. As an auditor, any peer should be able to evaluate a block or transaction in the same way that its PoV validators do during the validation.

Upon a misbehavior detection, the auditor generates a transaction with the evidence of the misbehavior in the contribution field (e.g., a soundness violating transaction in a committed block). The transaction then goes through the same PoV validation process as described earlier, except that the validators verify the correctness of the transaction as the correctness of the reported evidence. Once the transaction is validated and placed into a committed block to the ledger, the misbehaving peer is penalized by a misbehavior penalty fee. The misbehavior penalty fee is another constant parameter of \textit{\lightchain} that is application dependent, and is paid in the favor of the auditor peer who reports the misbehavior. Once a misbehavior is recorded for a peer on a committed block, its identifier is blacklisted. The blacklisted peers are isolated by the honest peers and hence can no longer participate in \textit{\lightchain}, i.e., any incoming message from the blacklisted peers is discarded by the honest peers. \\

Table \ref{lightchain:table_asymptotic} summarizes the asymptotic operational complexities of \textit{\lightchain}. 

\begin{table}
\centering
{
    \begin{tabular}{ |l|l|l|  }
    \hline
    Operation & Type & Complexity\\ 
    \hline
    Joining Skip Graph Overlay & Message & $O(\log{\systemCapacity})$ \\
    Randomized Bootstrapping & Message & $O(\log{\systemCapacity})$ \\ 
    TxB Generation/Validation & Time & $O(\log{\systemCapacity})$ \\
    TxB Generation/Validation & Message & $O(\log{\systemCapacity})$ \\
    TxB Storage & Storage & $O(\frac{\blockCapacity}{\systemCapacity})$ \\ 
    Direct State Retrieval & Message & $O(\log{\systemCapacity})$ \\ 
    \hline
    \end{tabular}
}
\caption{Summary of asymptotic operational complexities of \textit{\lightchain} in a system with $\systemCapacity$ nodes and $\blockCapacity$ blocks. TxB stands for "Transaction and Block".}
\label{lightchain:table_asymptotic}
\end{table}

\section{Security Analysis}
\label{lightchain:sec_security}
This section provides a summary of the security analysis of \textit{\lightchain} and skips details and intermediate steps due to the page limit. The interested readers are referred to the full version of our work \cite{hassanzadeh2019lightchain} for the detailed security analysis. 

\textbf{Definitions:} We analyze the security of \textit{\lightchain} from the lens of integrity, data availability, and service availability \cite{goodrich2011introduction}. We define the \textit{integrity} of \textit{\lightchain} as the property that the views of the peers towards the ledger are not being changed, except by committing a new block of validated transactions to the current tail of the ledger solely by the designated PoV validators of that block. We define the \textit{data availability} as the property that every validated transaction or committed block is being accessible in a timely fashion \cite{goodrich2011introduction}. We define the \textit{service availability} as the availability of the Consensus and View layer protocols. The Consensus layer service availability means that the honest nodes who follow the \textit{\lightchain} protocols should be able to find enough honest PoV validators, in expectation, to validate their transaction and block, as specified in Section \ref{lightchain:subsec_consensus}. The availability of the View layer means that the honest nodes should be able to find enough honest view introducers, in expectation, to bootstrap to the system, as specified in Section \ref{lightchain:view_layer}. Defining the service availability in expectation means that the honest nodes should be able to eventually bootstrap to the system and have their transactions and blocks eventually validated. 

\textbf{Adversarial and Honest Behavior:} We model all participating peers as probabilistic Turing machines, whose running time is polynomial in the security parameter of the system (i.e., $\lambda$). We define honest behavior as the one that follows \textit{\lightchain} protocols specified in Section \ref{lightchain:sec_solution}, and maintains its availability and accessibility in the network. On the contrary, the adversarial behavior is the one that deliberately deviates from the \textit{\lightchain} protocols at arbitrary points. Similar to BFT-based approaches \cite{pease1980reaching}, we assume the existence of a Sybil adversarial party \cite{douceur2002sybil} that adaptively takes control over at most a fraction $\adv$ of peers in the system. The adversarial party aims at compromising the integrity, data availability, or service availability of \textit{\lightchain} by orchestrating attacks through its controlled set of corrupted peers. In our analysis, to presume the worst-case scenario, we assume no churn and failure for corrupted peers under the control of the adversary. For the honest peers, on the other hand, we model the churn with a uniform failure probability of $\ufp$, under the crash-recover model \cite{tanenbaum2007distributed}, i.e., an honest node fails with a probability of $\ufp$ and recovers back online after a while.

\textbf{Assumptions:} Based on the above definitions we summarize the assumptions in the following corollaries. We assume a \textit{\lightchain} system with $\systemCapacity$ peers, a partially synchronous network, an authenticated routing mechanism (as defined in Section \ref{lightchain:sec_preliminaries}), where honest nodes follow a uniform failure probability of $\ufp$ under the crash-recover model. We assume the existence of the negligible probability $\epsilon(\lambda)$ and a probabilistic polynomial-time adversary that adaptively controls a fraction $\adv$ of all peers in the system.

\begin{corollary}
\label{lightchain:cor_integrity} 
\textbf{(Integrity and Data Availability):} Under the specified assumptions above, selecting the \textit{Validators Threshold} $\validatorThreshold$ and the \textit{Signatures Threshold} $\signatureThreshold$ in such a way that it satisfies the inequalities shown by Equations \ref{lightchain:eq_alpha} and \ref{lightchain:eq_tm}, respectively, yields in the success probability of the adversary on breaking the integrity or data availability of \textit{\lightchain} bounded by $\epsilon(\lambda)$.
\end{corollary}

\begin{align}
    \validatorThreshold &\geq \frac{(\sqrt{ \adv} + \sqrt{\adv \times (\psi^{-1}(1- \epsilon(\lambda)))^{2} + 4}))^{2}}{4(1-\adv)}
    \label{lightchain:eq_alpha} \\
     \signatureThreshold &\geq (\sqrt{\validatorThreshold \adv(1-\adv)} \times \psi^{-1}(1- \epsilon(\lambda))) + \validatorThreshold \adv + 1
     \label{lightchain:eq_tm}
\end{align}

% \begin{corollary}
% \label{lightchain:thm_data_availability}
% \textbf{(Data Availability):} Under the specified assumptions above, selecting the \textit{Validators Threshold} $\validatorThreshold$ and the \textit{Signatures Threshold} $\signatureThreshold$ as shown by Equations \ref{lightchain:eq_alpha} and \ref{lightchain:eq_tm}, respectively, yields in the data availability of \textit{\lightchain} is preserved with a very high probability in the security parameter of the system, i.e., the probability of at least $1- \epsilon(\lambda)$.
% \end{corollary}

\begin{corollary}
\label{lightchain:cor_service_availability}
\textbf{(Service Availability):} Under the specified assumptions above, setting values of the \textit{Validators Threshold} $\validatorThreshold$ and the \textit{Signatures Threshold} $\signatureThreshold$ in such a way that it satisfies the inequality shown by Equation \ref{lightchain:eq_efficiency}, yields in \textit{\lightchain} providing its service availability (where one replica per block is available) in expectation. 
\end{corollary}

\begin{equation}
    t \leq \frac{\validatorThreshold(1-\adv)(1- \ufp)}{\adv + (1 - f)(1 - \ufp)}
    \label{lightchain:eq_efficiency}
\end{equation}

\section{Experimental Results}
\label{lightchain:sec_results}
\subsection{Simulation Results}
%%%%%%%%%%%%%%%%%%%%%%%%%%%%%%%%%%%
\label{lightchain:subsec_simulation_result}
\textbf{Setup: } We implemented \textit{\lightchain} over the SkipSim simulator \cite{hassanzadeh2020skipsim}, where each node follows the Bitcoin churn trace with an expected online and offline duration of $10.6$ and $2.8$ hours, respectively \cite{imtiaz2019churn}, and generates a transaction per hour \cite{blockchaincharts, bitnodes}. We simulated for $100$ \textit{\lightchain} systems each $\systemCapacity=10,000$ peers. Each system was simulated for $48$ hours. 

\begin{figure*}
\centering
\begin{minipage}[c]{.28\textwidth}
\centering
    \scalebox{0.51}{\begin{tikzpicture}
\begin{axis}[
    ylabel={Adversarial Success Probability},
    xlabel={Signatures Threshold ($\signatureThreshold$) },
    xmin=0, xmax=40,
    ymin=0, ymax=1,
    xtick={0, 5, 10, 15, 20, 25, 30, 35, 40},
    ytick={0, 0.1, 0.2, 0.3, 0.4, 0.5, 0.6, 0.7, 0.8, 0.9, 1},
    legend style={
    %legend style={font=\fontsize{6}{5}\selectfont},
    %legend pos=inner north east, legend cell align = left
    }, 
    ymajorgrids=true,
    grid style=dashed,
]

    \addplot
    coordinates 
    {
        (1,  .81)
        (2, .47)
        (3, .19)
        (4, .05)
        (5, .01)
        (6, 10^-3)
        (7, 10^-4)
        (8, 0)
        (9, 0)
        (10,0)
    };
    \addlegendentry{$\adv = 0.16, \validatorThreshold = 10$}
    
    \addplot
    coordinates 
    {
        (5,  .86)
        (10, .11)
        (12, .017)
        (15, 10^-4)
        (18, 0)
        (19, 0)
        (20, 0)
    };
    \addlegendentry{$\adv = 0.33, \validatorThreshold = 20$}
    
    \addplot
    coordinates 
    {
        (15, .95)
        (20, .53)
        (25, .06)
        (30, 10^-5)
        (35, 0)
        (36, 0)
        (37, 0)
        (38, 0)
        (39, 0)
    };
    \addlegendentry{$\adv = 0.51, \validatorThreshold = 39$}

\end{axis}
\end{tikzpicture}}
    \caption*{(a) Integrity}
\end{minipage}
\begin{minipage}[c]{.32\textwidth}
\centering
    \scalebox{0.51}{\begin{tikzpicture}
\begin{axis}[
    ylabel={Adversarial Success Probability},
    xlabel={Signatures Threshold ($\signatureThreshold$) },
    xmin=0, xmax=100,
    ymin=0, ymax=1,
    xtick={0, 10, 20, 30, 40, 50, 60, 70, 80, 90, 100},
    ytick={0, 0.1, 0.2, 0.3, 0.4, 0.5, 0.6, 0.7, 0.8, 0.9, 1},
    legend style={
    %legend style={font=\fontsize{6}{5}\selectfont},
    %legend pos=inner north east, legend cell align = left
    }, 
    ymajorgrids=true,
    grid style=dashed,
]

    \addplot
    coordinates 
    {
        (1, .81)
        (4, .17)
        (8,  10^-4)
        (10, 10^-6)
        (11, 0)
        (12, 0)
        (14, 0)
    };
    \addlegendentry{$\adv = 0.16, \validatorThreshold = 14$}
    
    \addplot
    coordinates 
    {
        (30, .9)
        (50, .3)
        (60, 10^-2)
        (70, 10^-5)
        (80, 0)
        (90, 0)
        (100, 0)
    };
    \addlegendentry{$\adv = 0.33, \validatorThreshold = 144$}

\end{axis}
\end{tikzpicture}}
    \caption*{(b) Integrity and Service Availability}
\end{minipage}
\begin{minipage}[c]{.28\textwidth}
\centering
    \scalebox{0.51}{\begin{tikzpicture}
\begin{axis}[
    ylabel={Average Blocks Availability},
    xlabel={Signatures Threshold ($\signatureThreshold$) },
    xmin=0, xmax=100,
    ymin=0, ymax=85,
    xtick={0, 10, 20, 30, 40, 50, 60, 70, 80, 90, 100},
    ytick={0, 10, 20, 30, 40, 50, 60, 70, 80, 85},
    legend style={
    legend pos=north west, legend cell align = left
    }, 
    ymajorgrids=true,
    grid style=dashed,
]
    \addplot
    coordinates 
    {
        (1,  1.6)
        (4,  4.3)
        (8,  7.5)
        (10, 9.4)
        (11, 10.4)
        (12, 11.1)
    };
    \addlegendentry{$\adv = 0.16, \validatorThreshold = 14$}
    
    \addplot
    coordinates 
    {
        (50,  43)
        (60,  51)
        (70,  60)
        (80,  69)
        (90,  77)
        (100, 85)
    };
    \addlegendentry{$\adv = 0.33, \validatorThreshold = 144$}

\end{axis}
\end{tikzpicture}}
    \caption*{(c) Data Availability}
\end{minipage}
\caption{Performance evaluation of \textit{\lightchain} concerning its integrity, and service and data availability. The performance is reported as averages over $100$ \textit{\lightchain} systems each with $10,000$ nodes. Each system is simulated for $48$ hours under the real churn traces of Bitcoin \cite{imtiaz2019churn}.}
\label{lightchain:fig_simulation_results}
\end{figure*}

\begin{figure*}
\centering
\begin{minipage}[c]{.28\textwidth}
\centering
    \scalebox{0.51}{\begin{tikzpicture}
\begin{axis}[
    ylabel={Average Block Formation Time (s)},
    xlabel={Transactions Number},
    xmin=0, xmax=40,
    ymin=0, ymax=70,
    xtick={0, 10, 20, 40},
    ytick={0, 5, 10, 15, 20, 25, 30, 35, 40, 45, 50, 55, 60, 65, 70},
    legend style={
    legend pos=north west, legend cell align = left
    }, 
    ymajorgrids=true,
    grid style=dashed,
]
    \addplot
    coordinates 
    {
        (10,  1.8)
        (20,  3.6)
        (40,  6.3)
    };
    \addlegendentry{$\adv = 0.16, \validatorThreshold = 14$}
    
    \addplot
    coordinates 
    {
        (10,  15.58)
        (20,  32.98)
        (40,  66.86)
    };
    \addlegendentry{$\adv = 0.33, \validatorThreshold = 144$}
\end{axis}
\end{tikzpicture}}
    \caption*{(a) The block formation time versus the number of transactions in a block}
\end{minipage}
\begin{minipage}[c]{.32\textwidth}
\centering
    \scalebox{0.51}{\begin{tikzpicture}
\begin{axis}[
    ylabel={Average Block Validation Time (s)},
    xlabel={Number of Transactions},
    xmin=0, xmax=40,
    ymin=0, ymax=75,
    xtick={0, 10, 20, 40},
    ytick={0, 5, 10, 15, 20, 25, 30, 35, 40, 45, 50, 55, 60, 65, 70, 75},
    legend style={
    legend pos=north west, legend cell align = left
    }, 
    ymajorgrids=true,
    grid style=dashed,
]
    \addplot
    coordinates 
    {
        (10,  9)
        (20,  18)
        (40,  35)
    };
    \addlegendentry{$\adv = 0.16, \validatorThreshold = 14$}
    
    \addplot
    coordinates 
    {
        (10,  25)
        (20,  37)
        (40,  60)
    };
    \addlegendentry{$\adv = 0.33, \validatorThreshold = 144$}
\end{axis}
\end{tikzpicture}}
    \caption*{(b) The block validation time versus the number of transactions in a block}
\end{minipage}
\begin{minipage}[c]{.28\textwidth}
\centering
    \scalebox{0.51}{\begin{tikzpicture}
\begin{axis}[
    ylabel={Average Block Size (KB)},
    xlabel={Transactions Number},
    xmin=0, xmax=40,
    ymin=0, ymax=800,
    xtick={0, 10, 20, 40},
    ytick={0, 100, 200, 300, 400, 500, 600, 700, 800},
    legend style={
    legend pos=north west, legend cell align = left
    }, 
    ymajorgrids=true,
    grid style=dashed,
]
    \addplot
    coordinates 
    {
        (10,  31)
        (20,  59)
        (40,  115)
    };
    \addlegendentry{$\adv = 0.16, \validatorThreshold = 14$}
    
    \addplot
    coordinates 
    {
        (10,  208)
        (20,  387)
        (40,  743)
    };
    \addlegendentry{$\adv = 0.33, \validatorThreshold = 144$}
\end{axis}
\end{tikzpicture}}
    \caption*{(c) The block size versus the number of transactions in a block}
\end{minipage}
\caption{The performance of the proof-of-concept of \textit{\lightchain} deployment on the Google Cloud Platform with $1000$ nodes and one million transactions.}
\label{lightchain:fig_poc_results}
\end{figure*}

\textbf{Integrity:} Figure \ref{lightchain:fig_simulation_results}.a shows the success probability of the adversary that controls fraction $\adv$ of corrupted peers on compromising the integrity of the system as defined in Section \ref{lightchain:sec_security}. In this figure, $\adv = 0.16$ corresponds to the largest fraction of colluding hash power in the Bitcoin network \cite{cryptoeprint:2016:919}. Considering that each node in \textit{\lightchain} has a uniform chance of involvement in consensus, $\adv = 0.33$ is beyond the adversarial fraction threshold of BFT and PoS-based blockchains, e.g., Hyperledger \cite{androulaki2018hyperledger}. Likewise, $\adv = 0.51$ is beyond the adversarial power threshold of the PoW, e.g., Bitcoin \cite{nakamoto2008bitcoin}. Supported by Corollary \ref{lightchain:cor_integrity}, as shown by Figure \ref{lightchain:fig_simulation_results}.a, for each simulated adversarial fraction, there exists a certain value of $\validatorThreshold$ in which the success probability of adversarial peers in compromising the integrity of \textit{\lightchain} exponentially converges to zero with the growth of $\signatureThreshold$. 
%Figure \ref{lightchain:fig_simulation_results}.a. also supports Corollary \ref{lightchain:cor_integrity}, in which for $\adv = 0.16$ we obtain $\validatorThreshold \geq 10$ and $\signatureThreshold \geq 8$. Similarly, for $\adv = 0.33$, the theorem results in $\validatorThreshold \geq 20$ and $\signatureThreshold \geq 19$. Finally, for $\adv = 0.51$, the theorem results in $\validatorThreshold \geq 39$ and $\signatureThreshold \geq 37$. 
\textbf{This implies that the integrity of \textit{\lightchain} is preserved even when corrupted peers become the majority.} 

\textbf{Service Availability:} Figure \ref{lightchain:fig_simulation_results}.b presents the integrity aspect of \textit{\lightchain} when its service availability is preserved. The validator thresholds of Figure \ref{lightchain:fig_simulation_results}.b are obtained by applying both Corollaries \ref{lightchain:cor_integrity} and \ref{lightchain:cor_service_availability}. %Accordingly, we obtain $\validatorThreshold \geq 14$ and $\signatureThreshold \geq 11$ for $\adv = 0.16$, and $\validatorThreshold \geq 144$ and $\signatureThreshold \geq 80$ for $\adv = 0.33$. Corollaries \ref{lightchain:cor_service_availability} and \ref{lightchain:cor_service_availability}
These corollaries are infeasible to satisfy together for $\adv > 0.5$. Hence, although there is no adversarial bound for the integrity of \textit{\lightchain} alone, its integrity under service availability is preserved for the adversarial fractions less than $0.5$. This means that in contrast to the state-of-the-art blockchains such as Bitcoin \cite{nakamoto2008bitcoin}, Ethereum \cite{wood2014ethereum}, and Hyperledger \cite{androulaki2018hyperledger}, that are fully compromised once the adversarial fraction of nodes goes beyond their inherent threshold, our proposed \textit{\lightchain} system halts when the adversarial fraction of nodes goes beyond its configured operational parameters. This allows the blockchain community to notice the security risks and perform a decentralized bootstrapping of the system by discarding the nodes with suspicious behavior, which is left as future work. %We scope out the bootstrapping of \textit{\lightchain} after such a halt, and address it in our future works.  

\textbf{Data Availability: }Figure \ref{lightchain:fig_simulation_results}.c illustrates the average number of available replicas for each block in the system at each time slot over $48$ hours of the simulation as $\signatureThreshold$ grows. Each validated block is replicated $\signatureThreshold + 1$ times in the system, i.e., $\signatureThreshold$ validators as well as the owner itself. As shown by Figure \ref{lightchain:fig_simulation_results}.c, the average availability of the blocks increases linearly with respect to $\signatureThreshold$. With the uniform failure probability of $\ufp$, having $\signatureThreshold + 1$ replicas for a block results in $(\signatureThreshold + 1)\times(1-\ufp)$ available replicas in expectation. 
%Hence, although following Corollary \ref{lightchain:cor_integrity}, the data availability of \textit{\lightchain} as a security measure is preserved in expectation,
Hence, as shown by Figure \ref{lightchain:fig_simulation_results}.c choosing $\signatureThreshold \geq \frac{1}{1-q} - 1$ results in an expected availability of at least one replica at each time slot. 
In Figure \ref{lightchain:fig_simulation_results}.c, the number of replicas for $t = 1$ is about $1.6$ replicas on the average, and grows linearly as $\signatureThreshold$ increases. An identical behavior is observed for the transactions' availability. 

\subsection{Cloud Platform Deployment}
We implemented and deployed a proof-of-concept version of the \textit{\lightchain}  \cite{hassanzadeh2020containerized, hassanzadeh2020skip} with $1000$ nodes for the adversarial fractions of $0.16$ and $0.33$, and block sizes of $10$, $20$, and $40$ transactions on the Google Cloud Platform, where each node generates a transaction per second for a total of $1000$ transactions per node. Due to the quota limitations, we run each node on a low-power hardware setup with $2.2$ GHz of processing power and $3.9$ GB of memory. To avoid these low-power nodes crashing of a high degree of concurrency, and memory limitation, we serialized those operations of the nodes that require synchronized communication, i.e., resources getting blocked for a response or timeout.  
Figure \ref{lightchain:fig_poc_results}.a shows the block formation time, which is the time that it takes for a node to collect a certain number of newly validated transactions into a block, generate the block metadata as specified in Section \ref{lightchain:sec_solution}, and request the validation of the block. Figure \ref{lightchain:fig_poc_results}.b shows the average local block validation time at each validator. Finally, Figure \ref{lightchain:fig_poc_results}.c presents the block size in KB as the number of transactions grows. The difference between $\adv = 0.16$ and $\adv = 0.33$ in the slope of growth of the block formation time (i.e., Figure \ref{lightchain:fig_poc_results}.a) and block size (i.e., Figure \ref{lightchain:fig_poc_results}.c) is due to the $10$ times growth in the number of validators per block. A similar pattern is also observed in the transaction time (i.e., generation, validation, and insertion in the Skip Graph overlay), which is about $7$ and $82$ seconds for $\adv = 0.16$ and $\adv = 0.33$ on average, respectively. As mentioned earlier due to hardware limitations the operations like transaction and block generation are serialized in our implementation on the Google Cloud Platform. This difference is however not evident in the block validation process (i.e., Figure \ref{lightchain:fig_poc_results}.b) since verifying the validators signature is a lighter operation than the network-based operations (i.e., transactions and block generation) and is done with some degree of concurrency.
%At the proof-of-concept level, with a bare minimum implementation, serialized operations, and without any production-level optimizations, based on Figures \ref{lightchain:fig_poc_results}.a and \ref{lightchain:fig_poc_results}.b, \textit{\lightchain} generates an average of $1$ and $0.33$ Transactions Per Second (TPS) under the adversarial fractions of $\adv = 0.16$ and $\adv = 0.33$, respectively, which is slower than the production-grade transaction rate of blockchains like Ethereum with $15$ TPS \cite{schaffer2019performance}. 

\subsection{Comparison With Mainstreams}
\label{lightchain:sec_comparison_mainstream}
\textbf{Advantages:} As shown by our experimental results in Figure \ref{lightchain:fig_simulation_results}.a, \textit{\lightchain} preserves its integrity under the corrupted majority power of the system (e.g., $\adv = 0.51$), which is in contrast to all mainstream blockchains including Bitcoin \cite{nakamoto2008bitcoin} and Ethereum \cite{wood2014ethereum} that are fully compromised on integrity under this adversarial fraction. Also, as shown in Figure \ref{lightchain:fig_simulation_results}.b, \textit{\lightchain} preserves both its integrity and service availability under the corrupted power fraction of $\adv = 0.33$ of system, which is in contrast to the mainstream PoS- and BFT-based blockchains (e.g., Hyperledger \cite{androulaki2018hyperledger}). Considering our Google Cloud Platform deployment with $1,000$ nodes, one million transactions, $25,000$ blocks, and $\adv=0.16$, \textit{\lightchain} only has a storage overhead of about $30$ MB per node to maintain the transactions and blocks and keep up with the protocol. This is around $\storageGain$ times less storage compared to the mainstream blockchains such as Bitcoin and Ethereum that result in a storage overhead of about $2$ GB per node under the same setup. Also, in our cloud deployment, bootstrapping a new \textit{\lightchain} node to the system at the block height of $25,000$ via the randomized bootstrapping protocol takes around $28$ seconds on average. Considering the average block retrieval time of around $400$ ms in our deployment, \textit{\lightchain} is around $\bootstrapGain$ times faster on bootstrapping a new node than the mainstream blockchains such as Bitcoin and Ethereum, which take around $3$ hours by imposing a new node to download and process all $25,000$ generated blocks. Finally, under the mentioned setup, each \textit{\lightchain} node is involved in and rewarded for an average of $348$ block generation decision makings with a standard deviation of around $1.3$ blocks, which denotes a uniform distribution of the consensus involvement among the nodes. This is in contrast to the mainstream blockchains like Bitcoin where nodes are involved in the block generation decision making based on their influence in the system, i.e., hash power.  

\textbf{Shortcomings:} \textit{\lightchain}'s implementation is currently at the prototype level, which hinders back its transaction throughput to around a single transaction per second on average. This makes the transaction throughput of \textit{\lightchain} lower than the mainstream blockchains like the Bitcoin and Ethereum that are capable of processing around $4$ and $12$ transactions per second under their production-level implementation \cite{blockchaincharts}. As our future work, we plan to optimize the throughput of \textit{\lightchain} through a production-level implementation and software engineering-level optimizations.

\section{Conclusion}
\label{lightchain:sec_conclusion}
To improve the communication and storage scalability, as well as the decentralization of blockchain architectures, we proposed \textit{\lightchain}, which is a novel blockchain architecture that operates over a DHT overlay. 
\textit{\lightchain} provides addressable peers, blocks, and transactions within the network, which makes them efficiently accessible in an on-demand manner. Using \textit{\lightchain}, no peer is required to store the entire ledger. Rather, each peer replicates a random subset of the blocks and transactions, and answers other peer's queries on those. \textit{\lightchain} is a fair blockchain as it considers a uniform chance for all the participating peers to be involved in the consensus protocol regardless of their influence in the system (e.g., hashing power or stake). We analyzed \textit{\lightchain} both mathematically and experimentally regarding its security and performance. 

\section*{Acknowledgement}
The authors thank Ali Utkan Şahin, Nazir Nayal, Shadi Sameh Hamdan, and Mohammad Kefah Issa for their implementation, and  T\"{U}B\.{I}TAK (the Scientific and Technological Research Council of Turkey) for project 119E088.

\bibliographystyle{IEEEtran}
\bibliography{references}
\vskip -3\baselineskip plus -1fil
\begin{IEEEbiography}
    [{\includegraphics[width=1in,height=1.25in,clip,keepaspectratio]{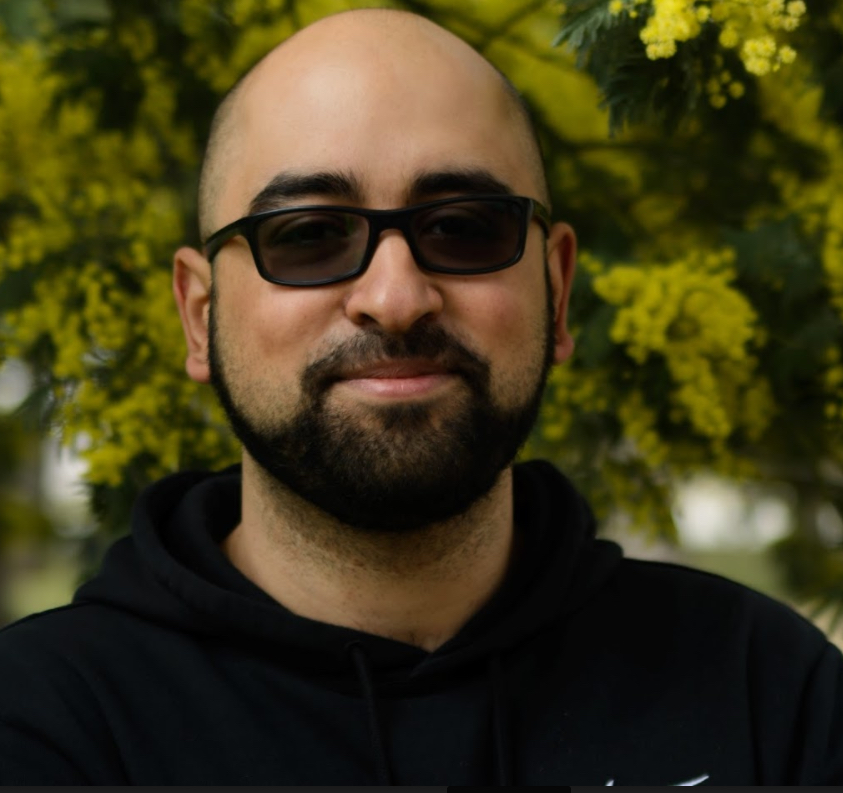}}]{Yahya Hassanzadeh-Nazarabadi} received his Ph.D. degree in Computer Science and Engineering from Koç University, Istanbul, Turkey, in 2019. His research interests are distributed cloud storage systems, blockchains, and security. Dr. Hassanzadeh is currently a Senior Distributed Systems Engineer at Dapper Labs, Vancouver, Canada.
\end{IEEEbiography}
\vskip -3\baselineskip plus -1fil
\begin{IEEEbiography}
    [{\includegraphics[width=1in,height=1.25in,clip,keepaspectratio]{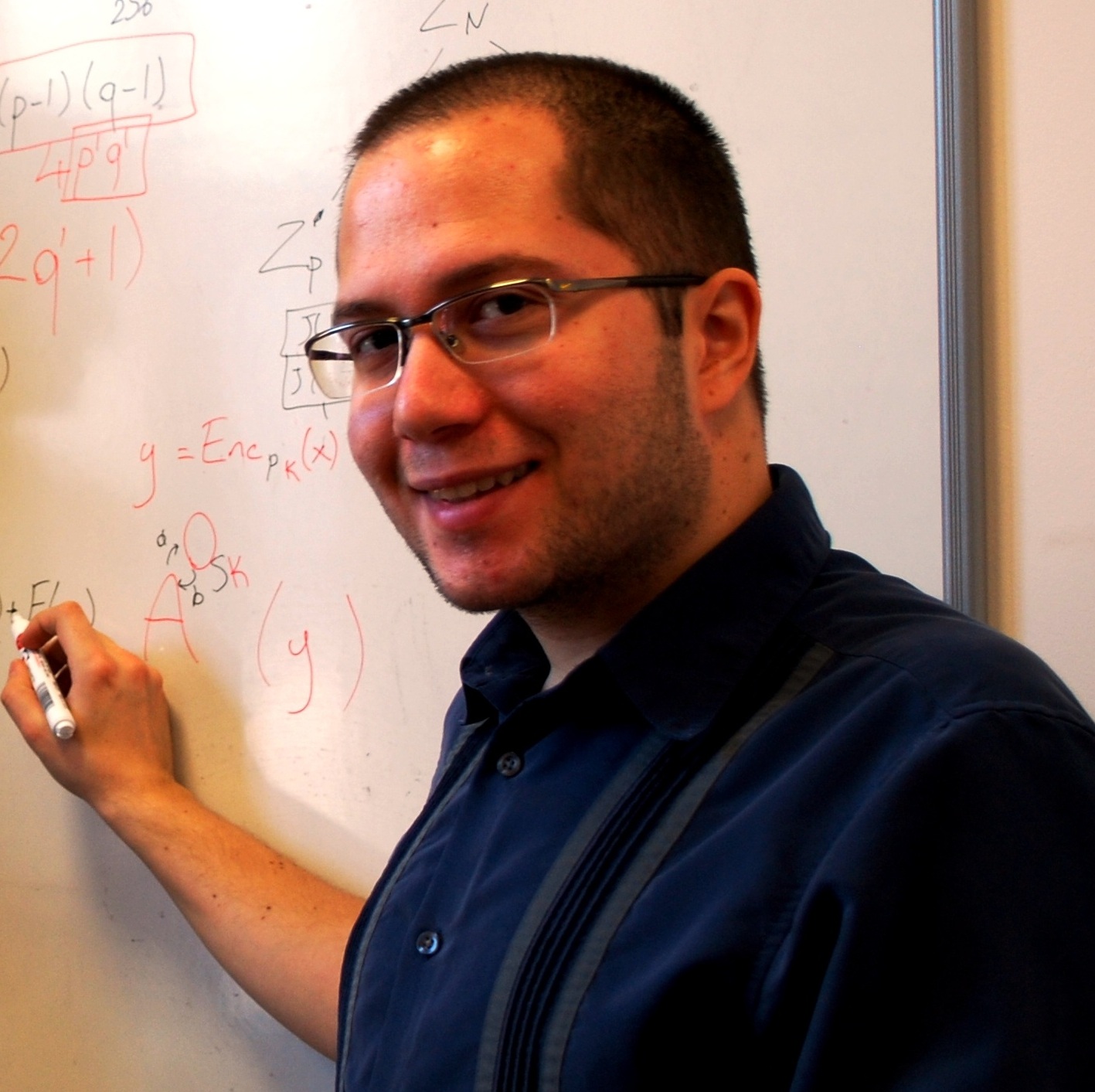}}]{Alptekin Küpçü} received his Ph.D. degree from Brown University Computer Science Department in 2010. Since then, he has been working as a faculty at Koç University, and leading the Cryptography, Cyber Security \& Privacy Research Group he has founded. Dr. Küpçü has various accomplishments including 6 international patents granted, 11 funded research projects (for 9 of which he was the principal investigator), 2 European Union COST Action management committee memberships, 4 Koç University Teaching Awards, 4 Outstanding Young Scientist Awards (BAGEP, GEBİP, METU Parlar Foundation, IEEE Turkey), ACM and IEEE Senior Member Awards, and the Royal Society of UK Newton Advanced Fellowship. For more information, visit https://crypto.ku.edu.tr
\end{IEEEbiography}
\vskip -2\baselineskip plus -1fil
\begin{IEEEbiography}
    [{\includegraphics[width=1in,height=1.25in,clip,keepaspectratio]{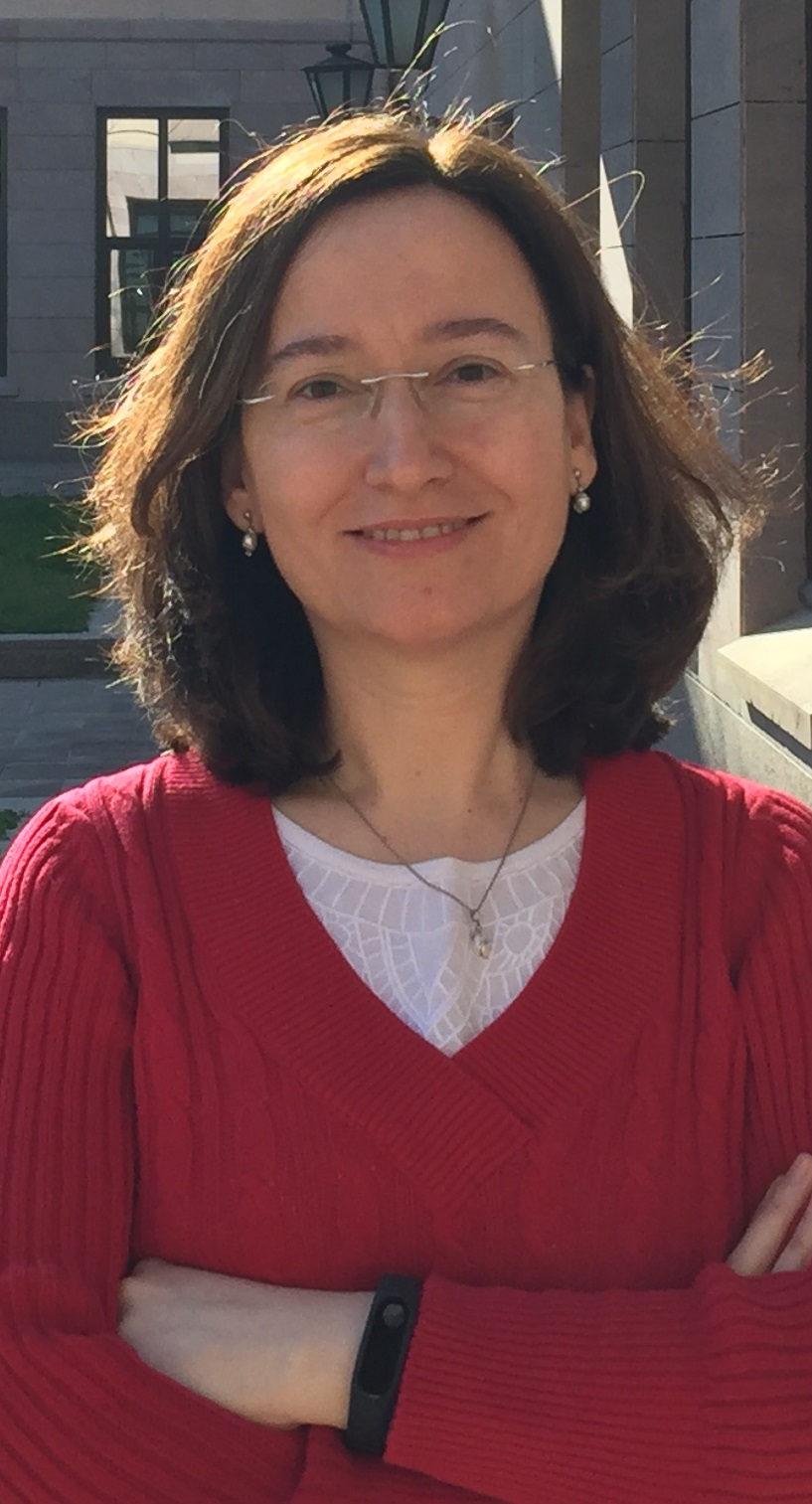}}]{Öznur Özkasap} received her Ph.D. degree in Computer Engineering from Ege University in 2000 and was a Graduate Research Assistant with the Department of Computer Science, Cornell University where she completed her Ph.D. dissertation. She is currently a Professor with the Department of Computer Engineering, Koç University, and leading the Distributed Systems and Reliable Networks (DISNET) Research Lab. She is the General Co-Chair of BCCA 2020 (International Conference on Blockchain Computing and Applications), and the Area Editor for Future Generation Computer Systems, Elsevier and Cluster Computing, Springer. She is a recipient of The Informatics Association of Turkey 2019 Prof. Aydın Köksal Computer Engineering Science Award.

\end{IEEEbiography}
\end{document}